\documentclass[aps,prb,twocolumn]{revtex4-2}
\usepackage{graphicx}
\usepackage{bm}
\usepackage{amsmath}
\usepackage{amssymb}
\usepackage{appendix}
\usepackage{subfigure}
\usepackage{xcolor, soul, color}
\sethlcolor{red}
\draft 

\begin{document}


\title{Intrinsic magnetism in KTaO$_3$ heterostructures} 

\author{Patrick W. Krantz}
\affiliation{Department of Physics, Northwestern University, Evanston, Illinois. 60208, USA}%
\author{Alexander Tyner}%
\affiliation{Graduate Program for Applied Physics, Northwestern University, Evanston, Illinois. 60208, USA}%
\author{Pallab Goswami}
\affiliation{Department of Physics, Northwestern University, Evanston, Illinois. 60208, USA}%
\affiliation{Graduate Program for Applied Physics, Northwestern University, Evanston, Illinois. 60208, USA}%
\author{Venkat Chandrasekhar}
\affiliation{Department of Physics, Northwestern University, Evanston, Illinois. 60208, USA}%
\affiliation{Graduate Program for Applied Physics, Northwestern University, Evanston, Illinois. 60208, USA}%

\date{\today}

\begin{abstract}
There has been intense recent interest in the two-dimensional electron gases (2DEGs) that form at the surfaces and interfaces of KTaO$_3$ (KTO), with the discovery of superconductivity at temperatures significantly higher than those of similar 2DEGs based on SrTiO$_3$ (STO).  Like STO heterostructures, these KTO 2DEGs are formed by depositing an overlayer on top of appropriately prepared KTO surfaces.  Some of these overlayers are magnetic, and the resulting 2DEGs show signatures of this magnetism, including hysteresis in the magnetoresistance (MR).  Here we show that KTO 2DEGs fabricated by depositing AlO$_x$ on top of KTO also show hysteretic MR, indicative of long range magnetic order, even though the samples nominally contain no intrinsic magnetic elements.  The hysteresis appears in both the transverse and longitudinal resistance in magnetic fields both perpendicular to and in the plane of the 2DEG.  The hysteretic MR has different characteristic fields and shapes for surfaces of different crystal orientations, and vanishes above a few Kelvin.  Density functional theory (DFT) calculations indicate that the magnetism likely arises from Ta$^{4+}$ local moments created in the presence of oxygen vacancies.

\end{abstract}

\maketitle

Two-dimensional electron gases (2DEGs) in complex oxide heterostructures have seen a great deal of interest in the past two decades due to the variety of phenomena they exhibit. The most studied of these 2DEGs are those formed at the interface and surface of SrTiO$_3$ (STO). \cite{gariglio_research_2016, huang_interface_2018, pai_physics_2018}  The phenomena seen in STO-based heterostructures include gate-tunable conductivity, \cite{thiel_tunable_2006, cen_nanoscale_2008} superconductivity, \cite{reyren_superconducting_2007} spin-orbit interactions,\cite{caviglia_tunable_2010,ben_shalom_tuning_2010} spin-polarized transport, \cite{lesne_highly_2016, song_observation_2017} quantum interference effects \cite{goswami_quantum_2016} and magnetism,\cite{brinkman_magnetic_2007,dikin_coexistence_2011,bert_direct_2011,li_coexistence_2011} among others. In the past few years, interest has turned to 2DEGs based on KTaO$_3$ (KTO)\cite{zou_latio_2015,bareille_two-dimensional_2015,wadehra_emergent_2021,gupta_ktao_2022}.  Bulk KTO is similar to STO in that is a band insulator with a large gap ($\sim$3.5 eV),\cite{wemple_transport_1965} and it has a large dielectric constant that grows with decreasing temperature but saturates at low temperatures, \cite{fujii_KTO_1976} signaling a transition to a quantum paraelectric phase. \cite{fujishita_2016} 

STO heterostructures fabricated with nominally non-magnetic components have been shown to exhibit long-range magnetic order, which typically manifests itself as a hysteresis in the magnetoresistance (MR),\cite{dikin_coexistence_2011} but also has been observed in torque magnetometry\cite{li_coexistence_2011} and scanning SQUID measurements.\cite{bert_direct_2011}  KTO structures with magnetic overlayers such as EuO also show hysteretic MR,\cite{zhang_high_mobility_2018} but this hysteresis cannot be separated from the magnetic properties of the overlayer itself.  More recently, a small anomalous Hall effect was observed in KTO heterostructures fabricated with a nonmagnetic TiO$_x$ overlayer which was attributed to magnetism arising from the presence of oxygen vacancies in the 2DEG.\cite{al-tawhid_oxygen_2022}  Here we report the observation of robust hysteretic MR in KTO 2DEGs fabricated with nonmagnetic AlO$_x$ overlayers.  The hysteretic MR shares many properties with that observed\cite{mehta_evidence_2012} in STO 2DEGs: It appears only at very low temperatures, it depends on the sweep rate of the magnetic field, and it can be suppressed by applying a relatively modest in-plane field. Density functional theory (DFT) calculations show that the presence of oxygen vacancies results in the donation of an electron to nonmagnetic Ta$^{5+}$ ions, converting them to magnetic Ta$^{4+}$ ions, which we speculate interact to give rise to long-range ferromagnetic order.   The increase in the temperature at which the hysteretic behavior is seen with the density of charge carriers suggests that the interaction between moments is mediated by the conduction electrons in the 2DEG.   

The devices in this study were Hall bars of length 600 $\mu$m and width 50 $\mu$m patterned by photolithography followed by successive deposition and oxidation of thin layers of Al on KTO substrates with three different surface crystal orientations: (001), (110) and (111). Two different sets of samples were fabricated, differentiated by substrate preparation prior to spinning photoresist and patterning.  Prior to spinning photoresist, the substrates from both sets of samples were cleaned using standard cleaning procedures.  The substrates in one set were then annealed as described by Tomar \textit{et al};\cite{tomar_realization_2018} the substrates in the second set did not undergo annealing.  Samples from the first set had higher charge concentrations and showed superconducting transitions at temperatures comparable to those found by other groups, while samples from the second set had lower charge concentrations and did not show superconducting transitions down to our lowest measurement temperatures ($\sim$ 25 mK).  Measurements of the longitudinal and transverse low-frequency differential resistance as a function of back gate voltage $V_g$, temperature $T$ and applied external field $H$ were performed using standard low frequency ac lock-in techniques in a dilution refrigerator equipped with a two-axis magnet, enabling MR measurements with the field aligned perpendicular to or in the plane of the 2DEG. While both sets of samples showed hysteretic behavior, we focus here on the data from this second set of samples, as they showed hysteretic behavior at low temperatures without the added complication of the superconducting response.  For these samples, one Hall bar was fabricated on a (001) oriented surface with the length of the Hall bar oriented along a $<001>$ surface crystal direction, and two others were fabricated on a (110) surface, with the lengths of the Hall bars oriented along the [001] and $[1\bar{1}0]$ direction respectively.  For the first set of samples that went superconducting, Hall bars were fabricated on (001), (110) and (111) surfaces, with all three orientations showing hysteretic behavior. 
 
\begin{figure}
\includegraphics[width=\columnwidth]{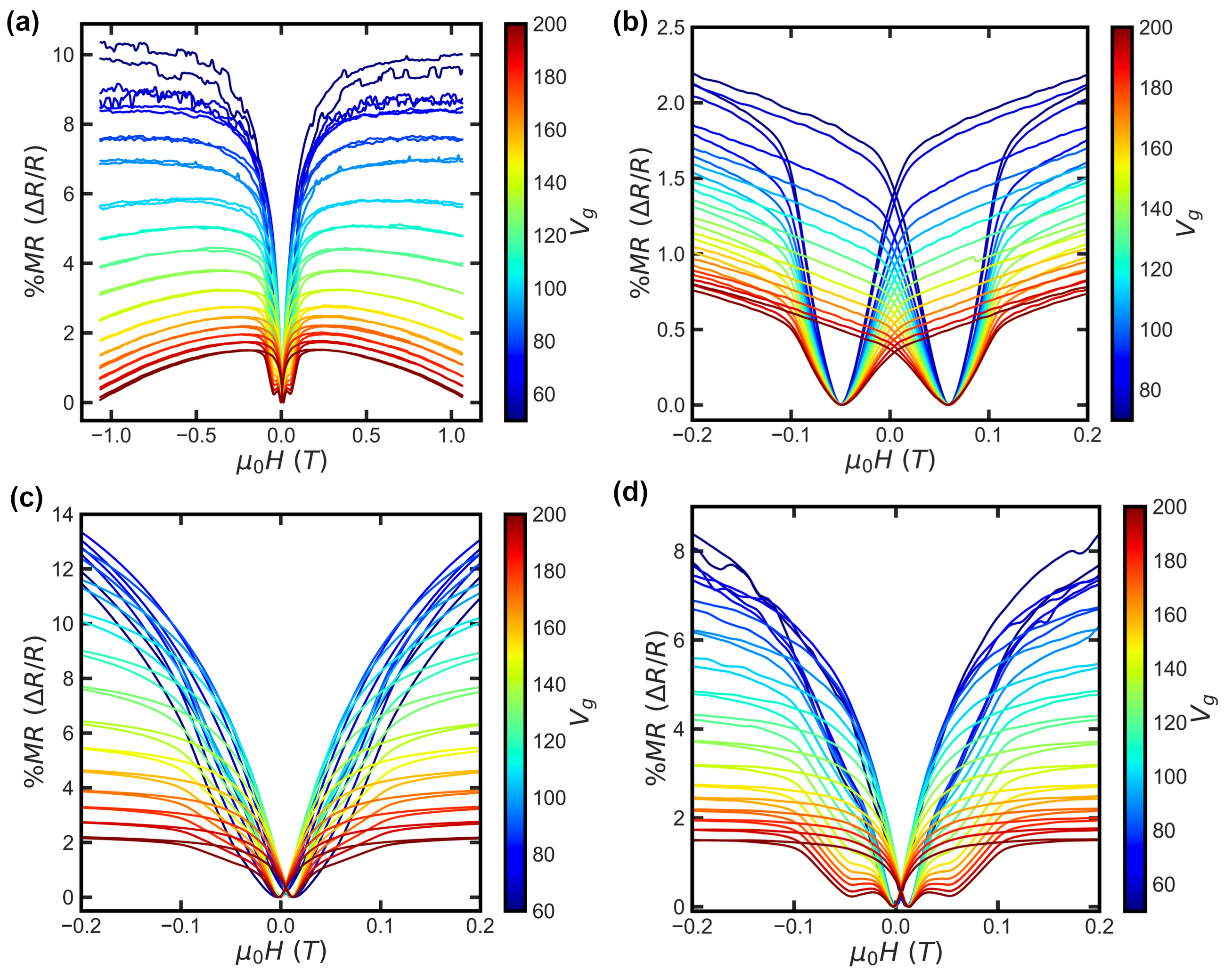}
\caption{  Longitudinal perpendicular field magnetoresistance at 30 mK at different gate voltages $V_g$.  (a) (110) sample with the Hall bar aligned along the [001] surface crystal direction over an expanded field range. (b-d)  Narrower field range data: (b)(001) device, (c) (110) device along the [100] direction, and (c) (110) device along the $[1\bar{1}0]$ direction (zoomed in view of (a)).}
\label{fig:longMR}
\end{figure} 

Figure \ref{fig:longMR}(a) shows the low temperature longitudinal MR of the (110) Hall bar aligned along the [001] surface crystal direction in a magnetic field applied perpendicular to the 2DEG at various back gate voltages $V_g$.  Before we discuss the hysteretic behavior, we consider the overall shape and magnitude of the MR.  A positive MR observed in similar complex oxide heterostructures is frequently ascribed to weak anti-localization,\cite{caviglia_tunable_2010} i.e., coherent quantum interference effects in the presence of strong spin-orbit interactions.  Since the spin-orbit interactions in KTO are more than an order of magnitude larger than in {\color{red}STO}, at first sight this is indeed a plausible origin of the MR.\cite{nakamura_electric_2009}  However, the \textit{maximum} amplitude of weak (anti)localization effects in 2D is $\sim R_{\square}/\pi R_Q$ to within factors of order unity, where $R_{\square}$ is the resistance per square of the 2DEG and $R_Q = h/e^2 \sim 26$ k$\Omega$ is the quantum of resistance.  $R_\square\sim 5$ k$\Omega$ for the sample whose data are shown in Fig. \ref{fig:longMR}(a) at the minimum $V_g$ of 60 V, so that the expected maximum MR due to quantum interference effects would be $\sim$6 \%, while the observed amplitude is almost 20 \%.  Again, this is the expected \textit{maximum} magnitude of the weak localization MR, i.e., the magnitude in the limit of a fully coherent sample.  In reality, the electron phase coherence length $L_\phi$ is quite short in such highly disordered samples.  If indeed the MR in Fig. \ref{fig:longMR}(a) were due to weak antilocalization, one can roughly estimate $L_\phi$ from the full width at half maximum magnetic field, $H_{FWHM}$, which is related to $L_\phi$ by $H_{FWHM}\sim \Phi_0/L_\phi^2$ ($\Phi_0=h/2e$ is the superconducting flux quantum).  With $H_{FWHM}\sim$ 0.1 T, one obtains $L_\phi \sim$ 150 nm, which as noted above, is quite short, and would be expected to substantially reduce the magnitude of the weak antilocalization MR.  Another possible origin for the large MR might be scattering from localized moments, since (as we show below) we do have underlying magnetism in the samples.  However, applying a field would be expected to align the moments and hence reduce the scattering, leading a decrease in resistance and a negative MR,\cite{sasaki_mr_1965} in contrast to the positive MR we observe.  At the moment, the origin of the large positive MR that we observe is not clear, although we speculate that it might be associated with the large spin-orbit interactions in KTO 2DEGs.

\begin{figure}
	\includegraphics[width=1.05\columnwidth]{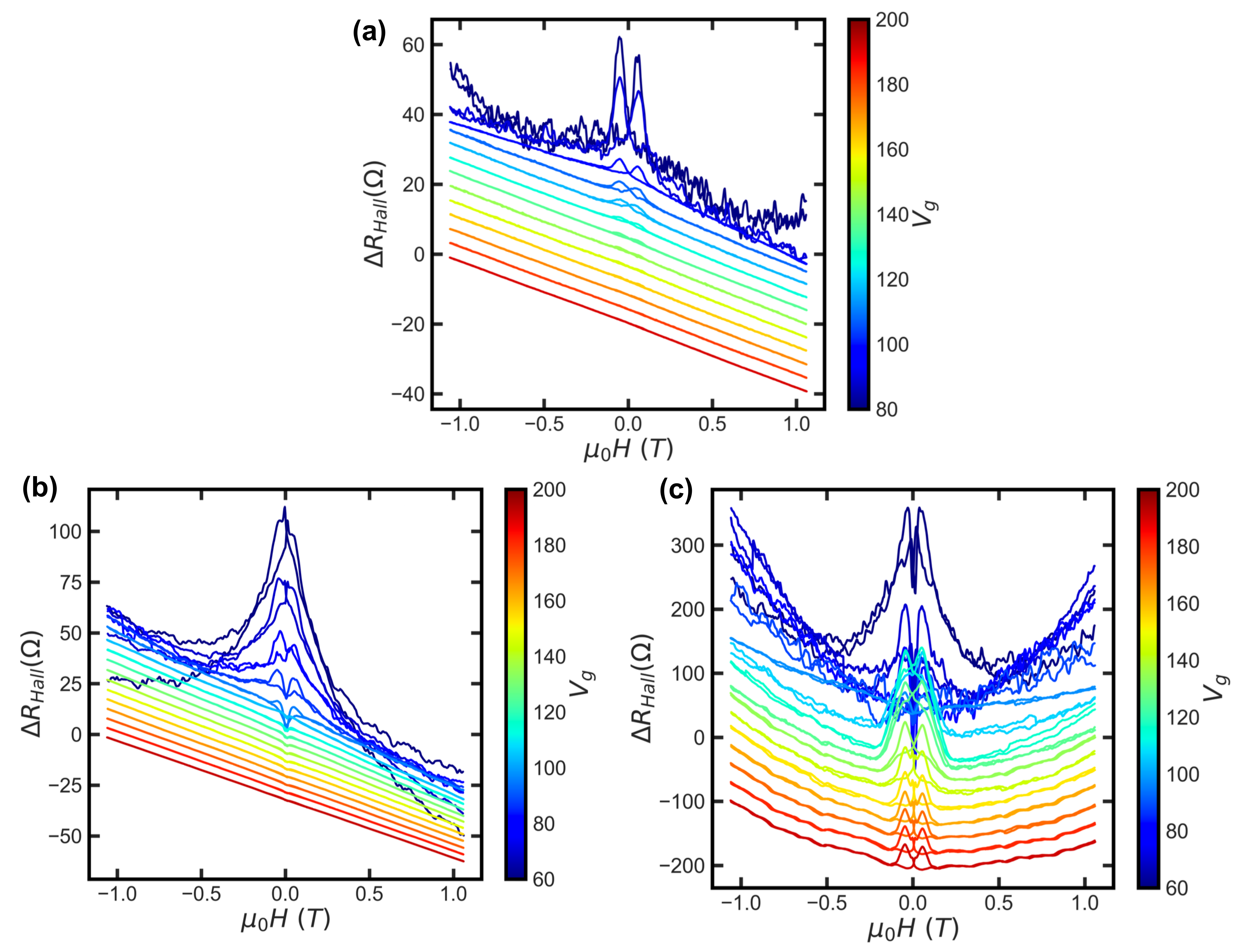}
	\caption{  Transverse (Hall) perpendicular field magnetoresistance at 30 mK at different gate voltages $V_g$.  (a) (001) oriented sample. (b) (110) oriented device along with the Hall bar along [100] direction, and (c) (110) oriented device with the Hall bar along $[1\bar{1}0]$ direction.  Curves have been shifted for clarity.}
	\label{fig:TransMR}
\end{figure}

We now consider the hysteresis in the MR.  Figures \ref{fig:longMR}(b-c) show the longitudinal MR for the (001) and two (110) Hall bars over a narrower field range to emphasize the hysteresis. The hysteresis manifests itself as minima in the MR. For the (001) oriented device, these minima occur at $\pm \sim 50$ mT (Fig. \ref{fig:longMR}(b)) and for the (110) device with the Hall bar oriented along the $[1\bar{1}0]$  direction, two pairs of minima are observed: one at $\pm 50$ mT and the other at $\pm 7$ mT.  For the (110) device with the Hall bar oriented along the $[100]$  direction, minima are observed at $\pm \sim 7$ mT, but there is also a hysteretic feature at $\pm 50$ mT, although this is not as well developed as for the $[1\bar{1}0]$ aligned Hall bar. (Note that there is small zero field offset in the data due to the remnant field of the superconducting solenoid.)  Figure \ref{fig:TransMR} shows similar data for the transverse (Hall) MR, taken simultaneously with the respective longitudinal component in Fig. \ref{fig:longMR}.  The transverse MR is also hysteretic, with hysteretic peaks instead of dips, but the peaks are only observed around $\pm 7$ mT:  no feature is observed around $\pm 50$ mT in any of the devices, as in the longitudinal MR of the (110) oriented devices.  More surprisingly, the hysteresis in the (001) oriented device and the (110) device with the Hall bar aligned along the [100] direction disappears as $V_g$ is increased, the MR being linear, and antisymmetric and non-hysteretic for $V_g \geq 160$ V, in sharp contrast to the corresponding longitudinal data shown in Fig. \ref{fig:longMR}.               

\begin{figure}
	\includegraphics[width=1.05\columnwidth]{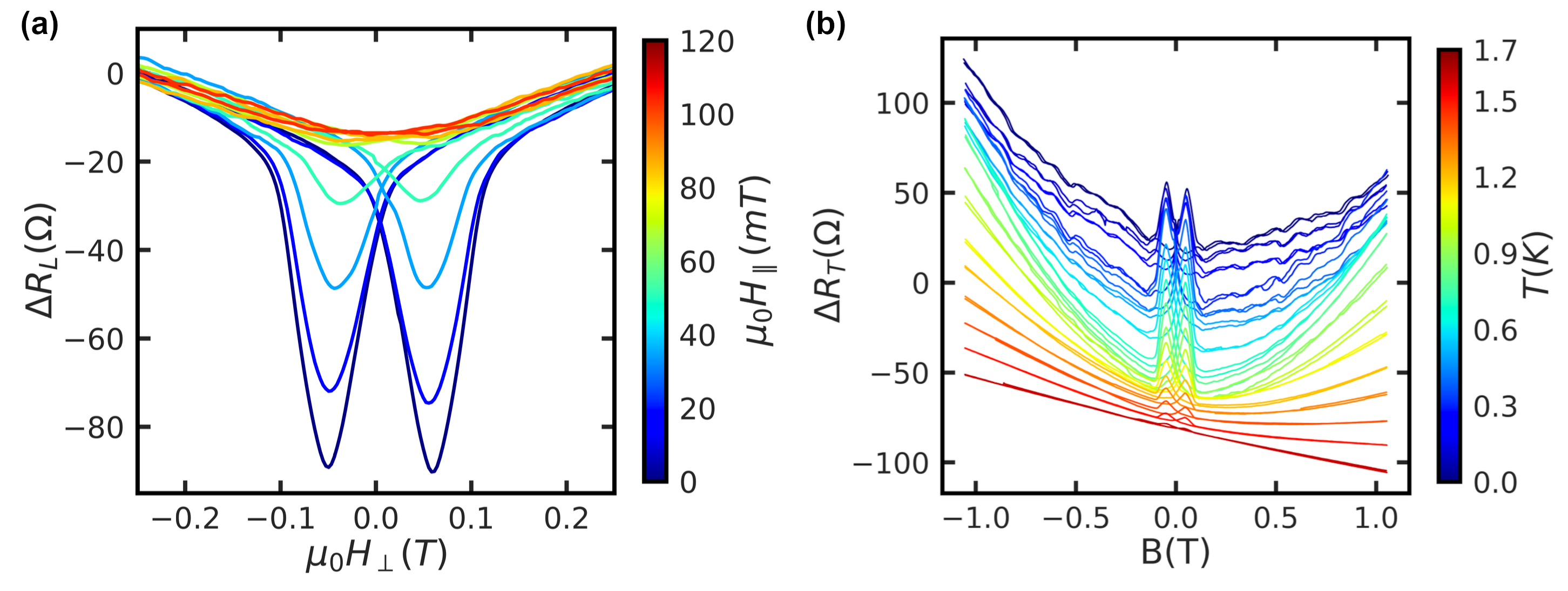}
	\caption{  (a) Longitudinal MR of the (001) oriented device at 30 mK as a function of magnetic field applied in the plane of the 2DEG.  (b) Temperature dependence of the transverse MR of the (110) oriented device with the Hall bar along the $[1\bar{1}0]$ direction.}
	\label{fig:ParallelFieldTempDep}
\end{figure}

Measuring MR for a magnetic field applied in the plane of the 2DEG provides evidence that the hysteresis is magnetic in origin. These data are shown in Fig. \ref{fig:ParallelFieldTempDep}(a), showing the longitudinal MR of the (001) oriented Hall bar.  The hysteresis progressively reduces in magnitude with an increasing magnetic field applied in the plane of the 2DEG, and completely disappears with a parallel field of $\sim$100 mT.  This is consistent with the picture that a field of this magnitude aligns the magnetization.  The hysteretic behavior also disappears with increasing temperature, as shown in Fig. \ref{fig:ParallelFieldTempDep}(b), which shows the transverse MR of the (110) device with the Hall bar aligned along the $[1\bar{1}0]$ direction. The hysteresis in the MR disappears by about 2 K.  

The materials used in the fabrication of our devices are all nominally non-magnetic, so it is at first surprising that we observe any signatures of magnetism at all.  Magnetism has also been reported in nominally non-magnetic STO heterostructures.\cite{dikin_coexistence_2011, bert_direct_2011,li_coexistence_2011,mehta_evidence_2012}  In that case, it was speculated that the magnetism arises from the ordering of Ti$^{3+}$ local magnetic moments that are formed in the presence of oxygen vacancies.\cite{pentcheva_charge_2006}  For KTO, our DFT calculations show that local Ta$^{4+}$ moments form in the presence of oxygen vacancies.  (Details of the DFT calculations can be found in the Supplementary.)  In general, the hysteretic behavior disappeared around this temperature for the second set of samples that did not go superconducting, while it persisted above $\sim$ 4 K for the first set of samples that went superconducting.  The key difference between these two sets of samples in terms of 2DEG properties was that the charge carrier concentration in the superconducting devices was about a factor of 2 larger than that in the normal devices.  This suggests that the long range magnetism in these devices may be mediated by the charge carriers.

Hysteretic MR associated with the reversal of the macroscopic magnetization under the influence of the external magnetic field is of course well-known in canonical ferromagnets such as Ni and Fe.\cite{aumentado_magnetoresistance_1999}  In the case of the canonical ferromagnets, the manifestation of the magnetism dynamics in the resistance comes through the mechanism of the so-called anisotropic magnetoresistance (AMR),\cite{nagaosa_anomalous_2010} which is associated with spin-orbit interactions and depends on the relative orientation between the magnetization and the direction of the measurement current.  The resistance is maximum when the magnetization is collinear with the current (either parallel or antiparallel) and minimum when it is perpendicular.\cite{mcguire_anisotropic_1975}  Even with such canonical ferromagnets, the MR depends on the magnetic crystalline anisotropy and the shape anisotropy of the magnetic structure, which dictate the direction along which the magnetization prefers to align, but more importantly on the direction of the magnetic field with respect to the current direction, as the magnetization will align with the magnetic field for large enough values of the magnetic field.  For example, if the external magnetic field direction is collinear with the current direction, then one would expect a maximum of the resistance for large (positive or negative) values of magnetic field where the magnetization is collinear with the current, and hysteretic minima at intermediate fields where the magnetization rotates in switching from one direction to the other.  On the other hand, if the magnetic field is applied perpendicular to the current direction, then one would expect to see a minimum at large values of magnetic field, and hysteretic maxima at intermediate fields where the magnetization switches.  This behavior has been observed in single domain Ni particles.\cite{aumentado_magnetoresistance_1999}  Since the magnetic field for our samples is applied perpendicular to the 2DEG and hence perpendicular to the current direction, we should expect to see hysteretic maxima in the MR if the origin is classic AMR, rather than the minima that we observe.

\begin{figure}
	\center{\includegraphics[width=0.7\columnwidth]{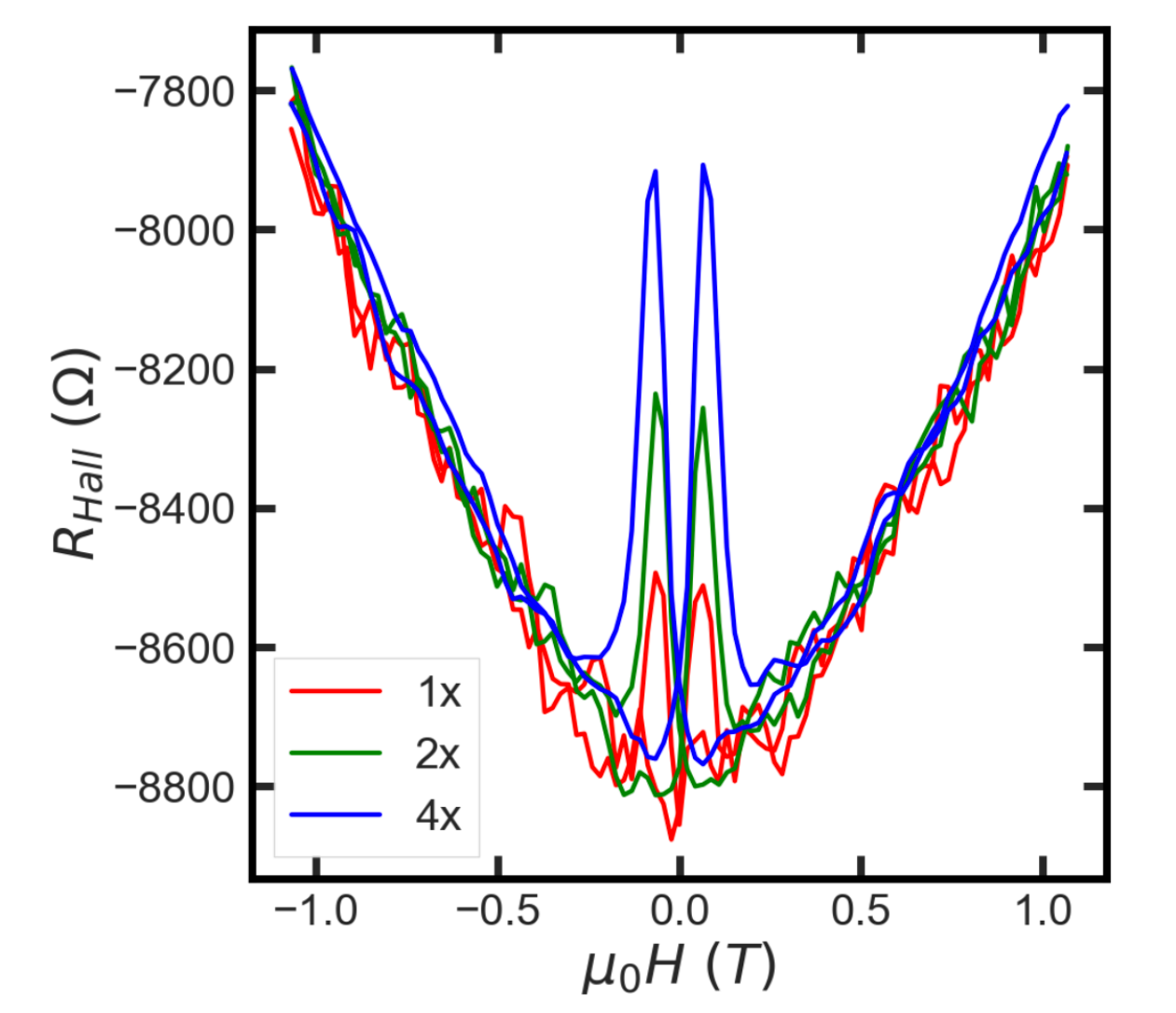}}
	\caption{  Transverse (Hall) MR of the (110) sample in the [1$\bar{1}$0] orientation at 30 mK for three different magnetic field sweep rates. }
	\label{fig:transratedep}
\end{figure}  

A further indication that the hysteresis in the MR that we observe is not due to classic AMR can be seen in the dependence of the magnitude of the hysteretic features on the sweep rate of the magnetic field.  Representative data are plotted in Fig. \ref{fig:transratedep}, which shows the transverse perpendicular field MR of the (110) device with the Hall bar oriented along the $[1\bar{1}0]$ direction at three different sweep rates.  The height of the hysteretic peaks increases with sweep rate. One does not expect the magnitude of the hysteretic features to depend on sweep rate if they arise from classic AMR.  If anything, one might expect the amplitude of the peaks to decrease if the time constants of the measurement electronics are not fast enough to capture rapid changes in the signal.  

A similar rate dependence of hysteretic MR features has been observed earlier in LaAlO$_3$(LAO)/STO heterostructures in the context of the superconductor to insulator transition.\cite{mehta_evidence_2012}  These devices show a co-existence of superconductivity and magnetism, and they can be tuned from the superconducting regime to the insulating regime by varying the backgate voltage $V_g$. In both the superconducting and insulating regime, the MR shows hysteresis.  In the insulating regime, the hysteresis manifests itself as dips in the longitudinal MR at the magnetic coercive field, similar to what is shown in Fig. \ref{fig:longMR} for the KTO devices.  In the superconducting regime, the hysteresis manifests itself as peaks in the longitudinal MR at the same coercive field values.  This transition from peaks to dips in the MR in going through the superconductor to insulator transition was analyzed in terms of charge-vortex duality.  In the superconducting regime, the rapidly changing magnetization reversal during the switching process results in vortex motion that leads to increased flux flow resistance, and hence resistance peaks.  The more rapidly the field is swept, the more rapid the motion of the vortices, the larger the flux flow resistance and hence the larger the peak height, as was observed.  On the insulating side, the idea is that charges are partially localized on conducting regions due to Coulomb interactions.  The rapidly changing magnetization during magnetization switching gives rise to large electric fields through Lenz's law, $\epsilon \sim \partial A/\partial t$, where $A$ is the vector potential arising from the local magnetic field.  The large electric fields allow charges localized on conducting regions to overcome the charging energy and hop between different conducting regions, lowering the resistance and leading to resistance dips.  The more rapidly the magnetization switches, the larger the electric fields and potentials and the greater the magnitude of the dips.  We believe that a similar mechanism gives rise to the hysteretic features in the KTO structures as well, although we do not have a wide enough range of magnetic sweep rates to verify the expected exponential dependence on the sweep rate.\cite{mehta_evidence_2012}      

In summary, KTO 2DEGs fabricated with nominally completely non-magnetic components show evidence of long-range magnetic order at low temperatures that manifests itself as hysteresis in the magnetoresistance.  The characteristics of the magnetoresistance in terms of the direction of the applied field as well as its dependence on the sweep rate of the magnetic field suggest that the origin of the hysteretic MR is not due to conventional anisotropic magnetoresistance, but arises from the interaction of partially localized charge carriers with the potentials created by the switching of the magnetization.  

This work was supported by the U.S. Department of Energy, Basic Energy Sciences, under Award No. DE-FG02-06ER46346. This work also made use of the NUFAB facility of Northwestern University’s NUANCE Center, which has received support from the SHyNE Resource (NSF ECCS-2025633), the IIN, and Northwestern’s MRSEC program (NSF DMR-1720139). Additional equipment support was provided by DURIP grant W911NF-20-1-0066.

\noindent\textbf{Data Availability Statement}  The data that support the findings of this study are available from
the corresponding author upon reasonable request.

\bibliography{KTO_bib}

\providecommand{\noopsort}[1]{}\providecommand{\singleletter}[1]{#1}%
\begin{thebibliography}{32}%
\makeatletter
\providecommand \@ifxundefined [1]{%
 \@ifx{#1\undefined}
}%
\providecommand \@ifnum [1]{%
 \ifnum #1\expandafter \@firstoftwo
 \else \expandafter \@secondoftwo
 \fi
}%
\providecommand \@ifx [1]{%
 \ifx #1\expandafter \@firstoftwo
 \else \expandafter \@secondoftwo
 \fi
}%
\providecommand \natexlab [1]{#1}%
\providecommand \enquote  [1]{``#1''}%
\providecommand \bibnamefont  [1]{#1}%
\providecommand \bibfnamefont [1]{#1}%
\providecommand \citenamefont [1]{#1}%
\providecommand \href@noop [0]{\@secondoftwo}%
\providecommand \href [0]{\begingroup \@sanitize@url \@href}%
\providecommand \@href[1]{\@@startlink{#1}\@@href}%
\providecommand \@@href[1]{\endgroup#1\@@endlink}%
\providecommand \@sanitize@url [0]{\catcode `\\12\catcode `\$12\catcode
  `\&12\catcode `\#12\catcode `\^12\catcode `\_12\catcode `\%12\relax}%
\providecommand \@@startlink[1]{}%
\providecommand \@@endlink[0]{}%
\providecommand \url  [0]{\begingroup\@sanitize@url \@url }%
\providecommand \@url [1]{\endgroup\@href {#1}{\urlprefix }}%
\providecommand \urlprefix  [0]{URL }%
\providecommand \Eprint [0]{\href }%
\providecommand \doibase [0]{http://dx.doi.org/}%
\providecommand \selectlanguage [0]{\@gobble}%
\providecommand \bibinfo  [0]{\@secondoftwo}%
\providecommand \bibfield  [0]{\@secondoftwo}%
\providecommand \translation [1]{[#1]}%
\providecommand \BibitemOpen [0]{}%
\providecommand \bibitemStop [0]{}%
\providecommand \bibitemNoStop [0]{.\EOS\space}%
\providecommand \EOS [0]{\spacefactor3000\relax}%
\providecommand \BibitemShut  [1]{\csname bibitem#1\endcsname}%
\let\auto@bib@innerbib\@empty
\bibitem [{\citenamefont {Gariglio}\ \emph {et~al.}(2016)\citenamefont
  {Gariglio}, \citenamefont {Gabay},\ and\ \citenamefont
  {Triscone}}]{gariglio_research_2016}%
  \BibitemOpen
  \bibfield  {author} {\bibinfo {author} {\bibfnamefont {S.}~\bibnamefont
  {Gariglio}}, \bibinfo {author} {\bibfnamefont {M.}~\bibnamefont {Gabay}}, \
  and\ \bibinfo {author} {\bibfnamefont {J.-M.}\ \bibnamefont {Triscone}},\
  }\href {\doibase 10.1063/1.4953822} {\bibfield  {journal} {\bibinfo
  {journal} {APL Materials}\ }\textbf {\bibinfo {volume} {4}},\ \bibinfo
  {pages} {060701} (\bibinfo {year} {2016})}\BibitemShut {NoStop}%
\bibitem [{\citenamefont {Huang}\ \emph {et~al.}(2018)\citenamefont {Huang},
  \citenamefont {{Ariando}}, \citenamefont {Renshaw~Wang}, \citenamefont
  {Rusydi}, \citenamefont {Chen}, \citenamefont {Yang},\ and\ \citenamefont
  {Venkatesan}}]{huang_interface_2018}%
  \BibitemOpen
  \bibfield  {author} {\bibinfo {author} {\bibfnamefont {Z.}~\bibnamefont
  {Huang}}, \bibinfo {author} {\bibnamefont {{Ariando}}}, \bibinfo {author}
  {\bibfnamefont {X.}~\bibnamefont {Renshaw~Wang}}, \bibinfo {author}
  {\bibfnamefont {A.}~\bibnamefont {Rusydi}}, \bibinfo {author} {\bibfnamefont
  {J.}~\bibnamefont {Chen}}, \bibinfo {author} {\bibfnamefont {H.}~\bibnamefont
  {Yang}}, \ and\ \bibinfo {author} {\bibfnamefont {T.}~\bibnamefont
  {Venkatesan}},\ }\href {\doibase 10.1002/adma.201802439} {\bibfield
  {journal} {\bibinfo  {journal} {Advanced Materials}\ }\textbf {\bibinfo
  {volume} {30}},\ \bibinfo {pages} {1802439} (\bibinfo {year}
  {2018})}\BibitemShut {NoStop}%
\bibitem [{\citenamefont {Pai}\ \emph {et~al.}(2018)\citenamefont {Pai},
  \citenamefont {Tylan-Tyler}, \citenamefont {Irvin},\ and\ \citenamefont
  {Levy}}]{pai_physics_2018}%
  \BibitemOpen
  \bibfield  {author} {\bibinfo {author} {\bibfnamefont {Y.-Y.}\ \bibnamefont
  {Pai}}, \bibinfo {author} {\bibfnamefont {A.}~\bibnamefont {Tylan-Tyler}},
  \bibinfo {author} {\bibfnamefont {P.}~\bibnamefont {Irvin}}, \ and\ \bibinfo
  {author} {\bibfnamefont {J.}~\bibnamefont {Levy}},\ }\href {\doibase
  10.1088/1361-6633/aa892d} {\bibfield  {journal} {\bibinfo  {journal} {Reports
  on Progress in Physics}\ }\textbf {\bibinfo {volume} {81}},\ \bibinfo {pages}
  {036503} (\bibinfo {year} {2018})}\BibitemShut {NoStop}%
\bibitem [{\citenamefont {Thiel}\ \emph {et~al.}(2006)\citenamefont {Thiel},
  \citenamefont {Hammerl}, \citenamefont {Schmehl}, \citenamefont {Schneider},\
  and\ \citenamefont {Mannhart}}]{thiel_tunable_2006}%
  \BibitemOpen
  \bibfield  {author} {\bibinfo {author} {\bibfnamefont {S.}~\bibnamefont
  {Thiel}}, \bibinfo {author} {\bibfnamefont {G.}~\bibnamefont {Hammerl}},
  \bibinfo {author} {\bibfnamefont {A.}~\bibnamefont {Schmehl}}, \bibinfo
  {author} {\bibfnamefont {C.~W.}\ \bibnamefont {Schneider}}, \ and\ \bibinfo
  {author} {\bibfnamefont {J.}~\bibnamefont {Mannhart}},\ }\href {\doibase
  10.1126/science.1131091} {\bibfield  {journal} {\bibinfo  {journal}
  {Science}\ }\textbf {\bibinfo {volume} {313}},\ \bibinfo {pages} {1942}
  (\bibinfo {year} {2006})}\BibitemShut {NoStop}%
\bibitem [{\citenamefont {Cen}\ \emph {et~al.}(2008)\citenamefont {Cen},
  \citenamefont {Thiel}, \citenamefont {Hammerl}, \citenamefont {Schneider},
  \citenamefont {Andersen}, \citenamefont {Hellberg}, \citenamefont
  {Mannhart},\ and\ \citenamefont {Levy}}]{cen_nanoscale_2008}%
  \BibitemOpen
  \bibfield  {author} {\bibinfo {author} {\bibfnamefont {C.}~\bibnamefont
  {Cen}}, \bibinfo {author} {\bibfnamefont {S.}~\bibnamefont {Thiel}}, \bibinfo
  {author} {\bibfnamefont {G.}~\bibnamefont {Hammerl}}, \bibinfo {author}
  {\bibfnamefont {C.~W.}\ \bibnamefont {Schneider}}, \bibinfo {author}
  {\bibfnamefont {K.~E.}\ \bibnamefont {Andersen}}, \bibinfo {author}
  {\bibfnamefont {C.~S.}\ \bibnamefont {Hellberg}}, \bibinfo {author}
  {\bibfnamefont {J.}~\bibnamefont {Mannhart}}, \ and\ \bibinfo {author}
  {\bibfnamefont {J.}~\bibnamefont {Levy}},\ }\href {\doibase 10.1038/nmat2136}
  {\bibfield  {journal} {\bibinfo  {journal} {Nature Materials}\ }\textbf
  {\bibinfo {volume} {7}},\ \bibinfo {pages} {298} (\bibinfo {year}
  {2008})}\BibitemShut {NoStop}%
\bibitem [{\citenamefont {Reyren}\ \emph {et~al.}(2007)\citenamefont {Reyren},
  \citenamefont {Thiel}, \citenamefont {Caviglia}, \citenamefont {Kourkoutis},
  \citenamefont {Hammerl}, \citenamefont {Richter}, \citenamefont {Schneider},
  \citenamefont {Kopp}, \citenamefont {Rüetschi}, \citenamefont {Jaccard},
  \citenamefont {Gabay}, \citenamefont {Muller}, \citenamefont {Triscone},\
  and\ \citenamefont {Mannhart}}]{reyren_superconducting_2007}%
  \BibitemOpen
  \bibfield  {author} {\bibinfo {author} {\bibfnamefont {N.}~\bibnamefont
  {Reyren}}, \bibinfo {author} {\bibfnamefont {S.}~\bibnamefont {Thiel}},
  \bibinfo {author} {\bibfnamefont {A.~D.}\ \bibnamefont {Caviglia}}, \bibinfo
  {author} {\bibfnamefont {L.~F.}\ \bibnamefont {Kourkoutis}}, \bibinfo
  {author} {\bibfnamefont {G.}~\bibnamefont {Hammerl}}, \bibinfo {author}
  {\bibfnamefont {C.}~\bibnamefont {Richter}}, \bibinfo {author} {\bibfnamefont
  {C.~W.}\ \bibnamefont {Schneider}}, \bibinfo {author} {\bibfnamefont
  {T.}~\bibnamefont {Kopp}}, \bibinfo {author} {\bibfnamefont {A.-S.}\
  \bibnamefont {Rüetschi}}, \bibinfo {author} {\bibfnamefont {D.}~\bibnamefont
  {Jaccard}}, \bibinfo {author} {\bibfnamefont {M.}~\bibnamefont {Gabay}},
  \bibinfo {author} {\bibfnamefont {D.~A.}\ \bibnamefont {Muller}}, \bibinfo
  {author} {\bibfnamefont {J.-M.}\ \bibnamefont {Triscone}}, \ and\ \bibinfo
  {author} {\bibfnamefont {J.}~\bibnamefont {Mannhart}},\ }\href {\doibase
  10.1126/science.1146006} {\bibfield  {journal} {\bibinfo  {journal}
  {Science}\ }\textbf {\bibinfo {volume} {317}},\ \bibinfo {pages} {1196}
  (\bibinfo {year} {2007})}\BibitemShut {NoStop}%
\bibitem [{\citenamefont {Caviglia}\ \emph {et~al.}(2010)\citenamefont
  {Caviglia}, \citenamefont {Gabay}, \citenamefont {Gariglio}, \citenamefont
  {Reyren}, \citenamefont {Cancellieri},\ and\ \citenamefont
  {Triscone}}]{caviglia_tunable_2010}%
  \BibitemOpen
  \bibfield  {author} {\bibinfo {author} {\bibfnamefont {A.~D.}\ \bibnamefont
  {Caviglia}}, \bibinfo {author} {\bibfnamefont {M.}~\bibnamefont {Gabay}},
  \bibinfo {author} {\bibfnamefont {S.}~\bibnamefont {Gariglio}}, \bibinfo
  {author} {\bibfnamefont {N.}~\bibnamefont {Reyren}}, \bibinfo {author}
  {\bibfnamefont {C.}~\bibnamefont {Cancellieri}}, \ and\ \bibinfo {author}
  {\bibfnamefont {J.-M.}\ \bibnamefont {Triscone}},\ }\href {\doibase
  10.1103/PhysRevLett.104.126803} {\bibfield  {journal} {\bibinfo  {journal}
  {Physical Review Letters}\ }\textbf {\bibinfo {volume} {104}},\ \bibinfo
  {pages} {126803} (\bibinfo {year} {2010})}\BibitemShut {NoStop}%
\bibitem [{\citenamefont {Ben~Shalom}\ \emph {et~al.}(2010)\citenamefont
  {Ben~Shalom}, \citenamefont {Sachs}, \citenamefont {Rakhmilevitch},
  \citenamefont {Palevski},\ and\ \citenamefont
  {Dagan}}]{ben_shalom_tuning_2010}%
  \BibitemOpen
  \bibfield  {author} {\bibinfo {author} {\bibfnamefont {M.}~\bibnamefont
  {Ben~Shalom}}, \bibinfo {author} {\bibfnamefont {M.}~\bibnamefont {Sachs}},
  \bibinfo {author} {\bibfnamefont {D.}~\bibnamefont {Rakhmilevitch}}, \bibinfo
  {author} {\bibfnamefont {A.}~\bibnamefont {Palevski}}, \ and\ \bibinfo
  {author} {\bibfnamefont {Y.}~\bibnamefont {Dagan}},\ }\href {\doibase
  10.1103/PhysRevLett.104.126802} {\bibfield  {journal} {\bibinfo  {journal}
  {Physical Review Letters}\ }\textbf {\bibinfo {volume} {104}},\ \bibinfo
  {pages} {126802} (\bibinfo {year} {2010})}\BibitemShut {NoStop}%
\bibitem [{\citenamefont {Lesne}\ \emph {et~al.}(2016)\citenamefont {Lesne},
  \citenamefont {Fu}, \citenamefont {Oyarzun}, \citenamefont {Rojas-Sánchez},
  \citenamefont {Vaz}, \citenamefont {Naganuma}, \citenamefont {Sicoli},
  \citenamefont {Attané}, \citenamefont {Jamet}, \citenamefont {Jacquet},
  \citenamefont {George}, \citenamefont {Barthélémy}, \citenamefont
  {Jaffrès}, \citenamefont {Fert}, \citenamefont {Bibes},\ and\ \citenamefont
  {Vila}}]{lesne_highly_2016}%
  \BibitemOpen
  \bibfield  {author} {\bibinfo {author} {\bibfnamefont {E.}~\bibnamefont
  {Lesne}}, \bibinfo {author} {\bibfnamefont {Y.}~\bibnamefont {Fu}}, \bibinfo
  {author} {\bibfnamefont {S.}~\bibnamefont {Oyarzun}}, \bibinfo {author}
  {\bibfnamefont {J.~C.}\ \bibnamefont {Rojas-Sánchez}}, \bibinfo {author}
  {\bibfnamefont {D.~C.}\ \bibnamefont {Vaz}}, \bibinfo {author} {\bibfnamefont
  {H.}~\bibnamefont {Naganuma}}, \bibinfo {author} {\bibfnamefont
  {G.}~\bibnamefont {Sicoli}}, \bibinfo {author} {\bibfnamefont {J.-P.}\
  \bibnamefont {Attané}}, \bibinfo {author} {\bibfnamefont {M.}~\bibnamefont
  {Jamet}}, \bibinfo {author} {\bibfnamefont {E.}~\bibnamefont {Jacquet}},
  \bibinfo {author} {\bibfnamefont {J.-M.}\ \bibnamefont {George}}, \bibinfo
  {author} {\bibfnamefont {A.}~\bibnamefont {Barthélémy}}, \bibinfo {author}
  {\bibfnamefont {H.}~\bibnamefont {Jaffrès}}, \bibinfo {author}
  {\bibfnamefont {A.}~\bibnamefont {Fert}}, \bibinfo {author} {\bibfnamefont
  {M.}~\bibnamefont {Bibes}}, \ and\ \bibinfo {author} {\bibfnamefont
  {L.}~\bibnamefont {Vila}},\ }\href {\doibase 10.1038/nmat4726} {\bibfield
  {journal} {\bibinfo  {journal} {Nature Materials}\ }\textbf {\bibinfo
  {volume} {15}},\ \bibinfo {pages} {1261} (\bibinfo {year}
  {2016})}\BibitemShut {NoStop}%
\bibitem [{\citenamefont {Song}\ \emph {et~al.}(2017)\citenamefont {Song},
  \citenamefont {Zhang}, \citenamefont {Su}, \citenamefont {Yuan},
  \citenamefont {Chen}, \citenamefont {Xing}, \citenamefont {Shi},
  \citenamefont {Sun},\ and\ \citenamefont {Han}}]{song_observation_2017}%
  \BibitemOpen
  \bibfield  {author} {\bibinfo {author} {\bibfnamefont {Q.}~\bibnamefont
  {Song}}, \bibinfo {author} {\bibfnamefont {H.}~\bibnamefont {Zhang}},
  \bibinfo {author} {\bibfnamefont {T.}~\bibnamefont {Su}}, \bibinfo {author}
  {\bibfnamefont {W.}~\bibnamefont {Yuan}}, \bibinfo {author} {\bibfnamefont
  {Y.}~\bibnamefont {Chen}}, \bibinfo {author} {\bibfnamefont {W.}~\bibnamefont
  {Xing}}, \bibinfo {author} {\bibfnamefont {J.}~\bibnamefont {Shi}}, \bibinfo
  {author} {\bibfnamefont {J.}~\bibnamefont {Sun}}, \ and\ \bibinfo {author}
  {\bibfnamefont {W.}~\bibnamefont {Han}},\ }\href {\doibase
  10.1126/sciadv.1602312} {\bibfield  {journal} {\bibinfo  {journal} {Science
  Advances}\ }\textbf {\bibinfo {volume} {3}},\ \bibinfo {pages} {e1602312}
  (\bibinfo {year} {2017})}\BibitemShut {NoStop}%
\bibitem [{\citenamefont {Goswami}\ \emph {et~al.}(2016)\citenamefont
  {Goswami}, \citenamefont {Mulazimoglu}, \citenamefont {Monteiro},
  \citenamefont {Wölbing}, \citenamefont {Koelle}, \citenamefont {Kleiner},
  \citenamefont {Blanter}, \citenamefont {Vandersypen},\ and\ \citenamefont
  {Caviglia}}]{goswami_quantum_2016}%
  \BibitemOpen
  \bibfield  {author} {\bibinfo {author} {\bibfnamefont {S.}~\bibnamefont
  {Goswami}}, \bibinfo {author} {\bibfnamefont {E.}~\bibnamefont
  {Mulazimoglu}}, \bibinfo {author} {\bibfnamefont {A.~M. R. V.~L.}\
  \bibnamefont {Monteiro}}, \bibinfo {author} {\bibfnamefont {R.}~\bibnamefont
  {Wölbing}}, \bibinfo {author} {\bibfnamefont {D.}~\bibnamefont {Koelle}},
  \bibinfo {author} {\bibfnamefont {R.}~\bibnamefont {Kleiner}}, \bibinfo
  {author} {\bibfnamefont {Y.~M.}\ \bibnamefont {Blanter}}, \bibinfo {author}
  {\bibfnamefont {L.~M.~K.}\ \bibnamefont {Vandersypen}}, \ and\ \bibinfo
  {author} {\bibfnamefont {A.~D.}\ \bibnamefont {Caviglia}},\ }\href {\doibase
  10.1038/nnano.2016.112} {\bibfield  {journal} {\bibinfo  {journal} {Nature
  Nanotechnology}\ }\textbf {\bibinfo {volume} {11}},\ \bibinfo {pages} {861}
  (\bibinfo {year} {2016})}\BibitemShut {NoStop}%
\bibitem [{\citenamefont {Brinkman}\ \emph {et~al.}(2007)\citenamefont
  {Brinkman}, \citenamefont {Huijben}, \citenamefont {van Zalk}, \citenamefont
  {Huijben}, \citenamefont {Zeitler}, \citenamefont {Maan}, \citenamefont
  {van~der Wiel}, \citenamefont {Rijnders}, \citenamefont {Blank},\ and\
  \citenamefont {Hilgenkamp}}]{brinkman_magnetic_2007}%
  \BibitemOpen
  \bibfield  {author} {\bibinfo {author} {\bibfnamefont {A.}~\bibnamefont
  {Brinkman}}, \bibinfo {author} {\bibfnamefont {M.}~\bibnamefont {Huijben}},
  \bibinfo {author} {\bibfnamefont {M.}~\bibnamefont {van Zalk}}, \bibinfo
  {author} {\bibfnamefont {J.}~\bibnamefont {Huijben}}, \bibinfo {author}
  {\bibfnamefont {U.}~\bibnamefont {Zeitler}}, \bibinfo {author} {\bibfnamefont
  {J.~C.}\ \bibnamefont {Maan}}, \bibinfo {author} {\bibfnamefont {W.~G.}\
  \bibnamefont {van~der Wiel}}, \bibinfo {author} {\bibfnamefont
  {G.}~\bibnamefont {Rijnders}}, \bibinfo {author} {\bibfnamefont {D.~H.~A.}\
  \bibnamefont {Blank}}, \ and\ \bibinfo {author} {\bibfnamefont
  {H.}~\bibnamefont {Hilgenkamp}},\ }\href {\doibase 10.1038/nmat1931}
  {\bibfield  {journal} {\bibinfo  {journal} {Nature Materials}\ }\textbf
  {\bibinfo {volume} {6}},\ \bibinfo {pages} {493} (\bibinfo {year}
  {2007})}\BibitemShut {NoStop}%
\bibitem [{\citenamefont {Dikin}\ \emph {et~al.}(2011)\citenamefont {Dikin},
  \citenamefont {Mehta}, \citenamefont {Bark}, \citenamefont {Folkman},
  \citenamefont {Eom},\ and\ \citenamefont
  {Chandrasekhar}}]{dikin_coexistence_2011}%
  \BibitemOpen
  \bibfield  {author} {\bibinfo {author} {\bibfnamefont {D.~A.}\ \bibnamefont
  {Dikin}}, \bibinfo {author} {\bibfnamefont {M.}~\bibnamefont {Mehta}},
  \bibinfo {author} {\bibfnamefont {C.~W.}\ \bibnamefont {Bark}}, \bibinfo
  {author} {\bibfnamefont {C.~M.}\ \bibnamefont {Folkman}}, \bibinfo {author}
  {\bibfnamefont {C.~B.}\ \bibnamefont {Eom}}, \ and\ \bibinfo {author}
  {\bibfnamefont {V.}~\bibnamefont {Chandrasekhar}},\ }\href {\doibase
  10.1103/PhysRevLett.107.056802} {\bibfield  {journal} {\bibinfo  {journal}
  {Physical Review Letters}\ }\textbf {\bibinfo {volume} {107}},\ \bibinfo
  {pages} {056802} (\bibinfo {year} {2011})}\BibitemShut {NoStop}%
\bibitem [{\citenamefont {Bert}\ \emph {et~al.}(2011)\citenamefont {Bert},
  \citenamefont {Kalisky}, \citenamefont {Bell}, \citenamefont {Kim},
  \citenamefont {Hikita}, \citenamefont {Hwang},\ and\ \citenamefont
  {Moler}}]{bert_direct_2011}%
  \BibitemOpen
  \bibfield  {author} {\bibinfo {author} {\bibfnamefont {J.~A.}\ \bibnamefont
  {Bert}}, \bibinfo {author} {\bibfnamefont {B.}~\bibnamefont {Kalisky}},
  \bibinfo {author} {\bibfnamefont {C.}~\bibnamefont {Bell}}, \bibinfo {author}
  {\bibfnamefont {M.}~\bibnamefont {Kim}}, \bibinfo {author} {\bibfnamefont
  {Y.}~\bibnamefont {Hikita}}, \bibinfo {author} {\bibfnamefont {H.~Y.}\
  \bibnamefont {Hwang}}, \ and\ \bibinfo {author} {\bibfnamefont {K.~A.}\
  \bibnamefont {Moler}},\ }\href {\doibase 10.1038/nphys2079} {\bibfield
  {journal} {\bibinfo  {journal} {Nature Physics}\ }\textbf {\bibinfo {volume}
  {7}},\ \bibinfo {pages} {767} (\bibinfo {year} {2011})}\BibitemShut {NoStop}%
\bibitem [{\citenamefont {Li}\ \emph {et~al.}(2011)\citenamefont {Li},
  \citenamefont {Richter}, \citenamefont {Mannhart},\ and\ \citenamefont
  {Ashoori}}]{li_coexistence_2011}%
  \BibitemOpen
  \bibfield  {author} {\bibinfo {author} {\bibfnamefont {L.}~\bibnamefont
  {Li}}, \bibinfo {author} {\bibfnamefont {C.}~\bibnamefont {Richter}},
  \bibinfo {author} {\bibfnamefont {J.}~\bibnamefont {Mannhart}}, \ and\
  \bibinfo {author} {\bibfnamefont {R.~C.}\ \bibnamefont {Ashoori}},\ }\href
  {\doibase 10.1038/nphys2080} {\bibfield  {journal} {\bibinfo  {journal}
  {Nature Physics}\ }\textbf {\bibinfo {volume} {7}},\ \bibinfo {pages} {762}
  (\bibinfo {year} {2011})}\BibitemShut {NoStop}%
\bibitem [{\citenamefont {Zou}\ \emph {et~al.}(2015)\citenamefont {Zou},
  \citenamefont {Ismail-Beigi}, \citenamefont {Kisslinger}, \citenamefont
  {Shen}, \citenamefont {Su}, \citenamefont {Walker},\ and\ \citenamefont
  {Ahn}}]{zou_latio_2015}%
  \BibitemOpen
  \bibfield  {author} {\bibinfo {author} {\bibfnamefont {K.}~\bibnamefont
  {Zou}}, \bibinfo {author} {\bibfnamefont {S.}~\bibnamefont {Ismail-Beigi}},
  \bibinfo {author} {\bibfnamefont {K.}~\bibnamefont {Kisslinger}}, \bibinfo
  {author} {\bibfnamefont {X.}~\bibnamefont {Shen}}, \bibinfo {author}
  {\bibfnamefont {D.}~\bibnamefont {Su}}, \bibinfo {author} {\bibfnamefont
  {F.~J.}\ \bibnamefont {Walker}}, \ and\ \bibinfo {author} {\bibfnamefont
  {C.~H.}\ \bibnamefont {Ahn}},\ }\href {\doibase 10.1063/1.4914310} {\bibfield
   {journal} {\bibinfo  {journal} {APL Materials}\ }\textbf {\bibinfo {volume}
  {3}},\ \bibinfo {pages} {036104} (\bibinfo {year} {2015})}\BibitemShut
  {NoStop}%
\bibitem [{\citenamefont {Bareille}\ \emph {et~al.}(2015)\citenamefont
  {Bareille}, \citenamefont {Fortuna}, \citenamefont {Rödel}, \citenamefont
  {Bertran}, \citenamefont {Gabay}, \citenamefont {Cubelos}, \citenamefont
  {Taleb-Ibrahimi}, \citenamefont {Le~Fèvre}, \citenamefont {Bibes},
  \citenamefont {Barthélémy}, \citenamefont {Maroutian}, \citenamefont
  {Lecoeur}, \citenamefont {Rozenberg},\ and\ \citenamefont
  {Santander-Syro}}]{bareille_two-dimensional_2015}%
  \BibitemOpen
  \bibfield  {author} {\bibinfo {author} {\bibfnamefont {C.}~\bibnamefont
  {Bareille}}, \bibinfo {author} {\bibfnamefont {F.}~\bibnamefont {Fortuna}},
  \bibinfo {author} {\bibfnamefont {T.~C.}\ \bibnamefont {Rödel}}, \bibinfo
  {author} {\bibfnamefont {F.}~\bibnamefont {Bertran}}, \bibinfo {author}
  {\bibfnamefont {M.}~\bibnamefont {Gabay}}, \bibinfo {author} {\bibfnamefont
  {O.~H.}\ \bibnamefont {Cubelos}}, \bibinfo {author} {\bibfnamefont
  {A.}~\bibnamefont {Taleb-Ibrahimi}}, \bibinfo {author} {\bibfnamefont
  {P.}~\bibnamefont {Le~Fèvre}}, \bibinfo {author} {\bibfnamefont
  {M.}~\bibnamefont {Bibes}}, \bibinfo {author} {\bibfnamefont
  {A.}~\bibnamefont {Barthélémy}}, \bibinfo {author} {\bibfnamefont
  {T.}~\bibnamefont {Maroutian}}, \bibinfo {author} {\bibfnamefont
  {P.}~\bibnamefont {Lecoeur}}, \bibinfo {author} {\bibfnamefont {M.~J.}\
  \bibnamefont {Rozenberg}}, \ and\ \bibinfo {author} {\bibfnamefont {A.~F.}\
  \bibnamefont {Santander-Syro}},\ }\href {\doibase 10.1038/srep03586}
  {\bibfield  {journal} {\bibinfo  {journal} {Scientific Reports}\ }\textbf
  {\bibinfo {volume} {4}},\ \bibinfo {pages} {3586} (\bibinfo {year}
  {2015})}\BibitemShut {NoStop}%
\bibitem [{\citenamefont {Wadehra}\ and\ \citenamefont
  {Chakraverty}(2021)}]{wadehra_emergent_2021}%
  \BibitemOpen
  \bibfield  {author} {\bibinfo {author} {\bibfnamefont {N.}~\bibnamefont
  {Wadehra}}\ and\ \bibinfo {author} {\bibfnamefont {S.}~\bibnamefont
  {Chakraverty}},\ }\href {\doibase 10.1007/s12034-021-02564-6} {\bibfield
  {journal} {\bibinfo  {journal} {Bulletin of Materials Science}\ }\textbf
  {\bibinfo {volume} {44}},\ \bibinfo {pages} {269} (\bibinfo {year}
  {2021})}\BibitemShut {NoStop}%
\bibitem [{\citenamefont {Gupta}\ \emph {et~al.}(2022)\citenamefont {Gupta},
  \citenamefont {Silotia}, \citenamefont {Kumari}, \citenamefont {Dumen},
  \citenamefont {Goyal}, \citenamefont {Tomar}, \citenamefont {Wadehra},
  \citenamefont {Ayyub},\ and\ \citenamefont {Chakraverty}}]{gupta_ktao_2022}%
  \BibitemOpen
  \bibfield  {author} {\bibinfo {author} {\bibfnamefont {A.}~\bibnamefont
  {Gupta}}, \bibinfo {author} {\bibfnamefont {H.}~\bibnamefont {Silotia}},
  \bibinfo {author} {\bibfnamefont {A.}~\bibnamefont {Kumari}}, \bibinfo
  {author} {\bibfnamefont {M.}~\bibnamefont {Dumen}}, \bibinfo {author}
  {\bibfnamefont {S.}~\bibnamefont {Goyal}}, \bibinfo {author} {\bibfnamefont
  {R.}~\bibnamefont {Tomar}}, \bibinfo {author} {\bibfnamefont
  {N.}~\bibnamefont {Wadehra}}, \bibinfo {author} {\bibfnamefont
  {P.}~\bibnamefont {Ayyub}}, \ and\ \bibinfo {author} {\bibfnamefont
  {S.}~\bibnamefont {Chakraverty}},\ }\href {\doibase 10.1002/adma.202106481}
  {\bibfield  {journal} {\bibinfo  {journal} {Advanced Materials}\ }\textbf
  {\bibinfo {volume} {34}},\ \bibinfo {pages} {2106481} (\bibinfo {year}
  {2022})}\BibitemShut {NoStop}%
\bibitem [{\citenamefont {Wemple}(1965)}]{wemple_transport_1965}%
  \BibitemOpen
  \bibfield  {author} {\bibinfo {author} {\bibfnamefont {S.~H.}\ \bibnamefont
  {Wemple}},\ }\href {\doibase 10.1103/PhysRev.137.A1575} {\bibfield  {journal}
  {\bibinfo  {journal} {Physical Review}\ }\textbf {\bibinfo {volume} {137}},\
  \bibinfo {pages} {A1575} (\bibinfo {year} {1965})}\BibitemShut {NoStop}%
\bibitem [{\citenamefont {Fujii}\ and\ \citenamefont
  {Sakudo}(1976)}]{fujii_KTO_1976}%
  \BibitemOpen
  \bibfield  {author} {\bibinfo {author} {\bibfnamefont {Y.}~\bibnamefont
  {Fujii}}\ and\ \bibinfo {author} {\bibfnamefont {T.}~\bibnamefont {Sakudo}},\
  }\href {\doibase 10.1143/JPSJ.41.888} {\bibfield  {journal} {\bibinfo
  {journal} {Journal of the Physical Society of Japan}\ }\textbf {\bibinfo
  {volume} {41}},\ \bibinfo {pages} {888} (\bibinfo {year} {1976})},\ \Eprint
  {http://arxiv.org/abs/https://doi.org/10.1143/JPSJ.41.888}
  {https://doi.org/10.1143/JPSJ.41.888} \BibitemShut {NoStop}%
\bibitem [{\citenamefont {Fujishita}\ \emph {et~al.}(2016)\citenamefont
  {Fujishita}, \citenamefont {Kitazawa}, \citenamefont {Saito}, \citenamefont
  {Ishisaka}, \citenamefont {Okamoto},\ and\ \citenamefont
  {Yamaguchi}}]{fujishita_2016}%
  \BibitemOpen
  \bibfield  {author} {\bibinfo {author} {\bibfnamefont {H.}~\bibnamefont
  {Fujishita}}, \bibinfo {author} {\bibfnamefont {S.}~\bibnamefont {Kitazawa}},
  \bibinfo {author} {\bibfnamefont {M.}~\bibnamefont {Saito}}, \bibinfo
  {author} {\bibfnamefont {R.}~\bibnamefont {Ishisaka}}, \bibinfo {author}
  {\bibfnamefont {H.}~\bibnamefont {Okamoto}}, \ and\ \bibinfo {author}
  {\bibfnamefont {T.}~\bibnamefont {Yamaguchi}},\ }\href {\doibase
  10.7566/JPSJ.85.074703} {\bibfield  {journal} {\bibinfo  {journal} {Journal
  of the Physical Society of Japan}\ }\textbf {\bibinfo {volume} {85}},\
  \bibinfo {pages} {074703} (\bibinfo {year} {2016})},\ \Eprint
  {http://arxiv.org/abs/https://doi.org/10.7566/JPSJ.85.074703}
  {https://doi.org/10.7566/JPSJ.85.074703} \BibitemShut {NoStop}%
\bibitem [{\citenamefont {Zhang}\ \emph {et~al.}(2018)\citenamefont {Zhang},
  \citenamefont {Yun}, \citenamefont {Zhang}, \citenamefont {Zhang},
  \citenamefont {Ma}, \citenamefont {Yan}, \citenamefont {Wang}, \citenamefont
  {Li}, \citenamefont {Li}, \citenamefont {Khan}, \citenamefont {Chen},
  \citenamefont {Liu}, \citenamefont {Hu}, \citenamefont {Liu}, \citenamefont
  {Shen}, \citenamefont {Han},\ and\ \citenamefont
  {Sun}}]{zhang_high_mobility_2018}%
  \BibitemOpen
  \bibfield  {author} {\bibinfo {author} {\bibfnamefont {H.}~\bibnamefont
  {Zhang}}, \bibinfo {author} {\bibfnamefont {Y.}~\bibnamefont {Yun}}, \bibinfo
  {author} {\bibfnamefont {X.}~\bibnamefont {Zhang}}, \bibinfo {author}
  {\bibfnamefont {H.}~\bibnamefont {Zhang}}, \bibinfo {author} {\bibfnamefont
  {Y.}~\bibnamefont {Ma}}, \bibinfo {author} {\bibfnamefont {X.}~\bibnamefont
  {Yan}}, \bibinfo {author} {\bibfnamefont {F.}~\bibnamefont {Wang}}, \bibinfo
  {author} {\bibfnamefont {G.}~\bibnamefont {Li}}, \bibinfo {author}
  {\bibfnamefont {R.}~\bibnamefont {Li}}, \bibinfo {author} {\bibfnamefont
  {T.}~\bibnamefont {Khan}}, \bibinfo {author} {\bibfnamefont {Y.}~\bibnamefont
  {Chen}}, \bibinfo {author} {\bibfnamefont {W.}~\bibnamefont {Liu}}, \bibinfo
  {author} {\bibfnamefont {F.}~\bibnamefont {Hu}}, \bibinfo {author}
  {\bibfnamefont {B.}~\bibnamefont {Liu}}, \bibinfo {author} {\bibfnamefont
  {B.}~\bibnamefont {Shen}}, \bibinfo {author} {\bibfnamefont {W.}~\bibnamefont
  {Han}}, \ and\ \bibinfo {author} {\bibfnamefont {J.}~\bibnamefont {Sun}},\
  }\href {\doibase 10.1103/PhysRevLett.121.116803} {\bibfield  {journal}
  {\bibinfo  {journal} {Physical Review Letters}\ }\textbf {\bibinfo {volume}
  {121}},\ \bibinfo {pages} {116803} (\bibinfo {year} {2018})}\BibitemShut
  {NoStop}%
\bibitem [{\citenamefont {Al-Tawhid}\ \emph {et~al.}(2022)\citenamefont
  {Al-Tawhid}, \citenamefont {Kanter}, \citenamefont {Hatefipour},
  \citenamefont {Irving}, \citenamefont {Kumah}, \citenamefont {Shabani},\ and\
  \citenamefont {Ahadi}}]{al-tawhid_oxygen_2022}%
  \BibitemOpen
  \bibfield  {author} {\bibinfo {author} {\bibfnamefont {A.~H.}\ \bibnamefont
  {Al-Tawhid}}, \bibinfo {author} {\bibfnamefont {J.}~\bibnamefont {Kanter}},
  \bibinfo {author} {\bibfnamefont {M.}~\bibnamefont {Hatefipour}}, \bibinfo
  {author} {\bibfnamefont {D.~L.}\ \bibnamefont {Irving}}, \bibinfo {author}
  {\bibfnamefont {D.~P.}\ \bibnamefont {Kumah}}, \bibinfo {author}
  {\bibfnamefont {J.}~\bibnamefont {Shabani}}, \ and\ \bibinfo {author}
  {\bibfnamefont {K.}~\bibnamefont {Ahadi}},\ }\href {\doibase
  10.1007/s11664-022-09941-9} {\bibfield  {journal} {\bibinfo  {journal} {J.
  Electron. Mater.}\ }\textbf {\bibinfo {volume} {51}},\ \bibinfo {pages}
  {7073} (\bibinfo {year} {2022})}\BibitemShut {NoStop}%
\bibitem [{\citenamefont {Mehta}\ \emph {et~al.}(2012)\citenamefont {Mehta},
  \citenamefont {Dikin}, \citenamefont {Bark}, \citenamefont {Ryu},
  \citenamefont {Folkman}, \citenamefont {Eom},\ and\ \citenamefont
  {Chandrasekhar}}]{mehta_evidence_2012}%
  \BibitemOpen
  \bibfield  {author} {\bibinfo {author} {\bibfnamefont {M.}~\bibnamefont
  {Mehta}}, \bibinfo {author} {\bibfnamefont {D.}~\bibnamefont {Dikin}},
  \bibinfo {author} {\bibfnamefont {C.}~\bibnamefont {Bark}}, \bibinfo {author}
  {\bibfnamefont {S.}~\bibnamefont {Ryu}}, \bibinfo {author} {\bibfnamefont
  {C.}~\bibnamefont {Folkman}}, \bibinfo {author} {\bibfnamefont
  {C.}~\bibnamefont {Eom}}, \ and\ \bibinfo {author} {\bibfnamefont
  {V.}~\bibnamefont {Chandrasekhar}},\ }\href {\doibase 10.1038/ncomms1959}
  {\bibfield  {journal} {\bibinfo  {journal} {Nature Communications}\ }\textbf
  {\bibinfo {volume} {3}},\ \bibinfo {pages} {955} (\bibinfo {year}
  {2012})}\BibitemShut {NoStop}%
\bibitem [{\citenamefont {Tomar}\ \emph {et~al.}(2018)\citenamefont {Tomar},
  \citenamefont {Wadehra}, \citenamefont {Budhiraja}, \citenamefont {Prakash},\
  and\ \citenamefont {Chakraverty}}]{tomar_realization_2018}%
  \BibitemOpen
  \bibfield  {author} {\bibinfo {author} {\bibfnamefont {R.}~\bibnamefont
  {Tomar}}, \bibinfo {author} {\bibfnamefont {N.}~\bibnamefont {Wadehra}},
  \bibinfo {author} {\bibfnamefont {V.}~\bibnamefont {Budhiraja}}, \bibinfo
  {author} {\bibfnamefont {B.}~\bibnamefont {Prakash}}, \ and\ \bibinfo
  {author} {\bibfnamefont {S.}~\bibnamefont {Chakraverty}},\ }\href {\doibase
  10.1016/j.apsusc.2017.08.101} {\bibfield  {journal} {\bibinfo  {journal}
  {Applied Surface Science}\ }\textbf {\bibinfo {volume} {427}},\ \bibinfo
  {pages} {861} (\bibinfo {year} {2018})}\BibitemShut {NoStop}%
\bibitem [{\citenamefont {Nakamura}\ and\ \citenamefont
  {Kimura}(2009)}]{nakamura_electric_2009}%
  \BibitemOpen
  \bibfield  {author} {\bibinfo {author} {\bibfnamefont {H.}~\bibnamefont
  {Nakamura}}\ and\ \bibinfo {author} {\bibfnamefont {T.}~\bibnamefont
  {Kimura}},\ }\href {\doibase 10.1103/PhysRevB.80.121308} {\bibfield
  {journal} {\bibinfo  {journal} {Physical Review B}\ }\textbf {\bibinfo
  {volume} {80}},\ \bibinfo {pages} {121308} (\bibinfo {year}
  {2009})}\BibitemShut {NoStop}%
\bibitem [{\citenamefont {Sasaki}(1965)}]{sasaki_mr_1965}%
  \BibitemOpen
  \bibfield  {author} {\bibinfo {author} {\bibfnamefont {W.}~\bibnamefont
  {Sasaki}},\ }\href {\doibase 10.1038/ncomms1959} {\bibfield  {journal}
  {\bibinfo  {journal} {Journal of the Physical Society of Japan}\ }\textbf
  {\bibinfo {volume} {20}},\ \bibinfo {pages} {825} (\bibinfo {year}
  {1965})}\BibitemShut {NoStop}%
\bibitem [{\citenamefont {Pentcheva}\ and\ \citenamefont
  {Pickett}(2006)}]{pentcheva_charge_2006}%
  \BibitemOpen
  \bibfield  {author} {\bibinfo {author} {\bibfnamefont {R.}~\bibnamefont
  {Pentcheva}}\ and\ \bibinfo {author} {\bibfnamefont {W.~E.}\ \bibnamefont
  {Pickett}},\ }\href {\doibase 10.1103/PhysRevB.74.035112} {\bibfield
  {journal} {\bibinfo  {journal} {Phys. Rev. B}\ }\textbf {\bibinfo {volume}
  {74}},\ \bibinfo {pages} {035112} (\bibinfo {year} {2006})}\BibitemShut
  {NoStop}%
\bibitem [{\citenamefont {Aumentado}\ and\ \citenamefont
  {Chandrasekhar}(1999)}]{aumentado_magnetoresistance_1999}%
  \BibitemOpen
  \bibfield  {author} {\bibinfo {author} {\bibfnamefont {J.}~\bibnamefont
  {Aumentado}}\ and\ \bibinfo {author} {\bibfnamefont {V.}~\bibnamefont
  {Chandrasekhar}},\ }\href {\doibase 10.1063/1.123706} {\bibfield  {journal}
  {\bibinfo  {journal} {Applied Physics Letters}\ }\textbf {\bibinfo {volume}
  {74}},\ \bibinfo {pages} {1898} (\bibinfo {year} {1999})},\ \bibinfo {note}
  {\_eprint:
  https://pubs.aip.org/aip/apl/article-pdf/74/13/1898/10179843/1898\_1\_online.pdf}\BibitemShut
  {NoStop}%
\bibitem [{\citenamefont {Nagaosa}\ \emph {et~al.}(2010)\citenamefont
  {Nagaosa}, \citenamefont {Sinova}, \citenamefont {Onoda}, \citenamefont
  {MacDonald},\ and\ \citenamefont {Ong}}]{nagaosa_anomalous_2010}%
  \BibitemOpen
  \bibfield  {author} {\bibinfo {author} {\bibfnamefont {N.}~\bibnamefont
  {Nagaosa}}, \bibinfo {author} {\bibfnamefont {J.}~\bibnamefont {Sinova}},
  \bibinfo {author} {\bibfnamefont {S.}~\bibnamefont {Onoda}}, \bibinfo
  {author} {\bibfnamefont {A.~H.}\ \bibnamefont {MacDonald}}, \ and\ \bibinfo
  {author} {\bibfnamefont {N.~P.}\ \bibnamefont {Ong}},\ }\href {\doibase
  10.1103/RevModPhys.82.1539} {\bibfield  {journal} {\bibinfo  {journal}
  {Reviews of Modern Physics}\ }\textbf {\bibinfo {volume} {82}},\ \bibinfo
  {pages} {1539} (\bibinfo {year} {2010})}\BibitemShut {NoStop}%
\bibitem [{\citenamefont {McGuire}\ and\ \citenamefont
  {Potter}(1975)}]{mcguire_anisotropic_1975}%
  \BibitemOpen
  \bibfield  {author} {\bibinfo {author} {\bibfnamefont {T.}~\bibnamefont
  {McGuire}}\ and\ \bibinfo {author} {\bibfnamefont {R.}~\bibnamefont
  {Potter}},\ }\href {\doibase 10.1109/TMAG.1975.1058782} {\bibfield  {journal}
  {\bibinfo  {journal} {IEEE Transactions on Magnetics}\ }\textbf {\bibinfo
  {volume} {11}},\ \bibinfo {pages} {1018} (\bibinfo {year}
  {1975})}\BibitemShut {NoStop}%
\end{thebibliography}%
\end{document}


\preprint{APS/123-QED}

\title{Supplementary Materials: Intrinsic magnetism in KTaO$_3$ heterostructures} 

\author{Patrick W. Krantz}
\affiliation{Department of Physics, Northwestern University, Evanston, Illinois. 60208, USA}%
\author{Alexander Tyner}%
\affiliation{Graduate Program for Applied Physics, Northwestern University, Evanston, Illinois. 60208, USA}%
\author{Pallab Goswami}
\affiliation{Department of Physics, Northwestern University, Evanston, Illinois. 60208, USA}%
\affiliation{Graduate Program for Applied Physics, Northwestern University, Evanston, Illinois. 60208, USA}%
\author{Venkat Chandrasekhar}
\affiliation{Department of Physics, Northwestern University, Evanston, Illinois. 60208, USA}%
\affiliation{Graduate Program for Applied Physics, Northwestern University, Evanston, Illinois. 60208, USA}%

\date{\today}

\maketitle

\section*{Computational details}
KTaO$_{3}$ belongs to spacegroup 221 with lattice parameter $a=4.03$ \AA. The lattice parameters and atomic positions are taken from the Materials Project \cite{jain_commentary_2013}. All first principles calculations are based on density functional theory and were performed using the Quantum ESPRESSO software package \cite{giannozzi_q_2020}. The calculations utilize the generalized gradient approximations (GGA) of Perdew-Burke-Ernzerhoff (PBE) \cite{perdew_generalized_1996}. Spin-orbit coupling is included in the calculations.

\paragraph*{Emergent Magnetism: The (001) Surface}
To examine the presence of emergent surface magnetism, leading to a finite zero field charge Hall effect, a supercell is constructed of size $2 \times 1 \times 6$. The (001) surface is perpendicular to the $c$-axis of the supercell and a minimum of 20 \AA of vacuum is added along the $c$-direction. We first focus on the (001) surface as it displays a zero field charge Hall conductivity while limiting computational expense. The terminations of the supercell are made symmetric. Multiple relaxations are performed, one for each oxygen vacancy that can be formed at the top layer. The relaxation calculations are performed using a plane-wave cutoff of 60 Ry, a charge cutoff of 550 Ry and the cell is sampled with a Monkhorst k-mesh of $2 \times 4 \times 1$, magnetization is considered in relaxation calculations and the bottom three layers are held fixed. Following relaxation the lowest energy structure is selected for a self consistent calculation. The self-consistent calculation is performed utilizing a Monkhorst k-mesh of $4 \times 8 \times 1$ followed by a non-self consistent calculation with a Monkhorst k-mesh of $8 \times 16 \times 1$. Following this, the density of states was calculated to determine spin-polarization and charge distribution. The simulation estimates a total magnetic moment of $~0.56$ Bohr magneton for the full supercell with two surface Ta atoms. This result is consistent with the relative magnetude of the experimental results for the (001) sample. The process was repeated with no oxygen vacancies to ensure the system converged to a non-magnetic state. 
\begin{figure*}
\centering
\subfigure[]{
\includegraphics[scale=0.4]{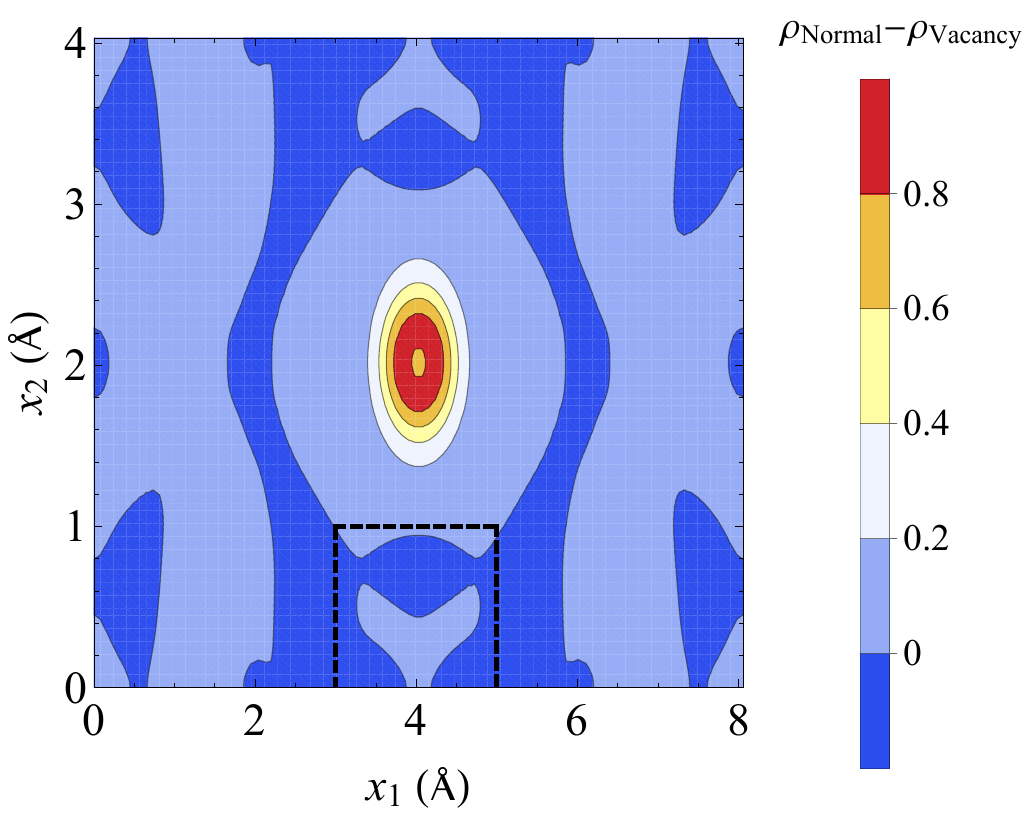}
\label{fig:CD1001}}
\subfigure[]{
\includegraphics[scale=0.4]{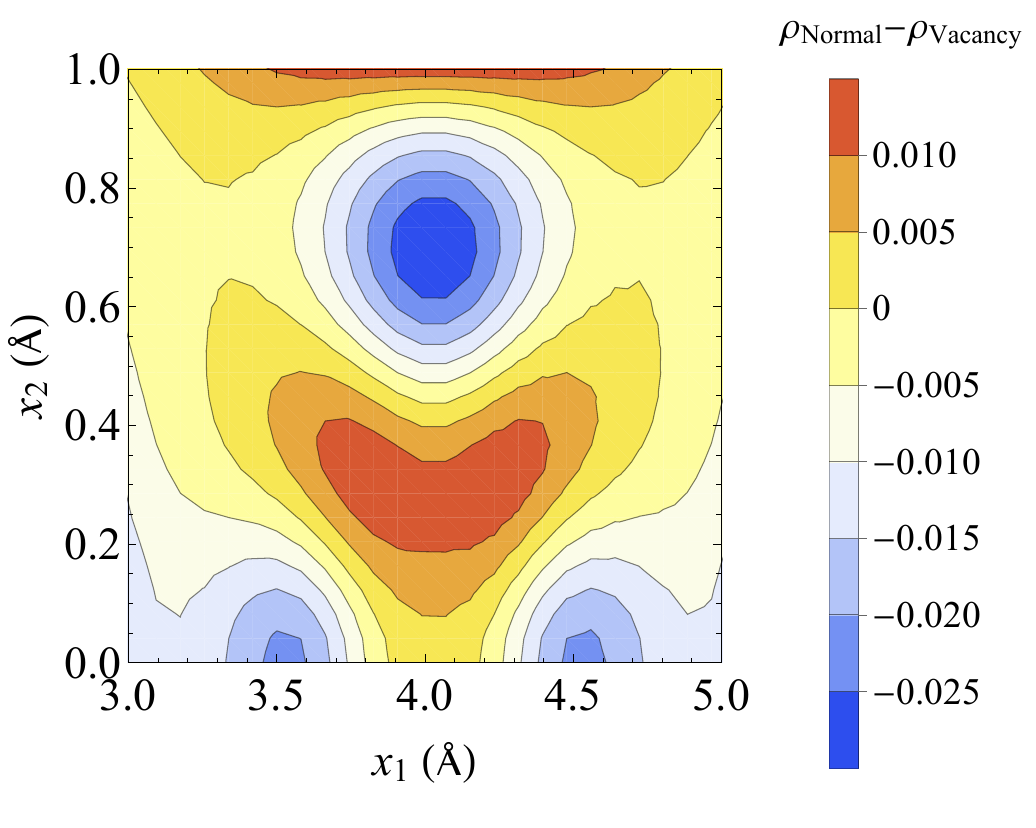}
\label{fig:CD2001}}
\subfigure[]{
\includegraphics[scale=0.4]{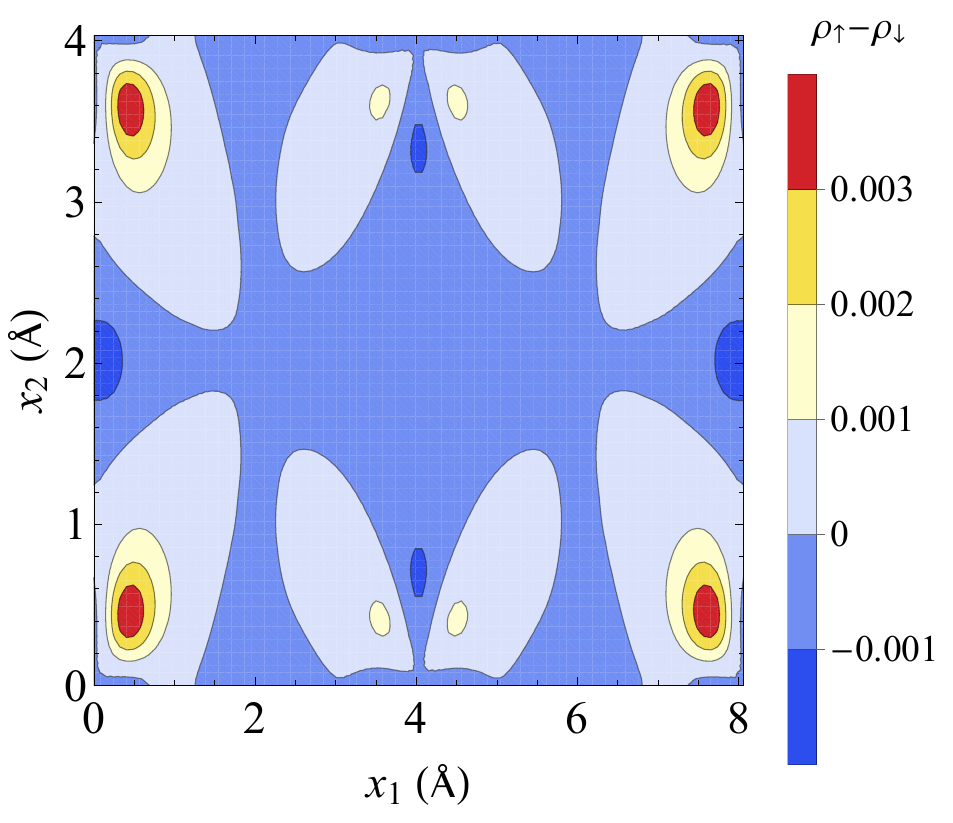}
\label{fig:SP001}}
\caption{(a) Difference in charge density on the surface for the supercell with an oxygen vacancy at the center ($\rho_{vacancy}$) and without ($\rho_{normal}$). The dashed box outlines the location of the Ta atom nearest to the vacancy. The location of the oxygen vacancy at the center is evident by the increased values in that region. (b) The charge density inside the dashed box shown in (a). (c) Spin polarization on the surface of the supercell with an oxygen vacancy. Any significant contribution to spin-polarization is localized to the Ta atoms.}
\label{001SurfaceDFT}
\end{figure*}

The partial charges of each atom were then compared for the systems with and without the charge vacancy. The atom which displays most significant change in partial charge is the Ta atom closest to the oxygen vacancy on the surface with a change in charge of approximately $0.5 e^{-}$. This is visible in the density plot shown in Fig. \ref{001SurfaceDFT}(a).   \newline{}

\begin{figure*}[h!]
\centering
\subfigure[]{
\includegraphics[scale=0.6]{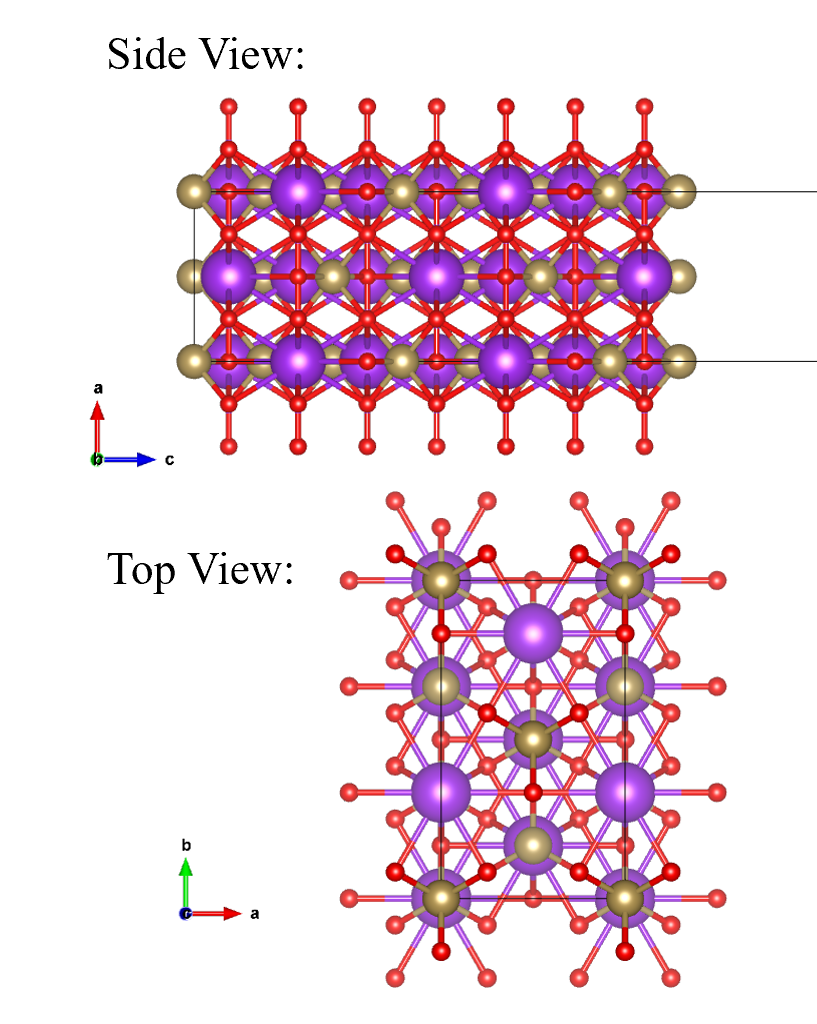}
\label{fig:111_structure}}
\subfigure[]{
\includegraphics[scale=0.4]{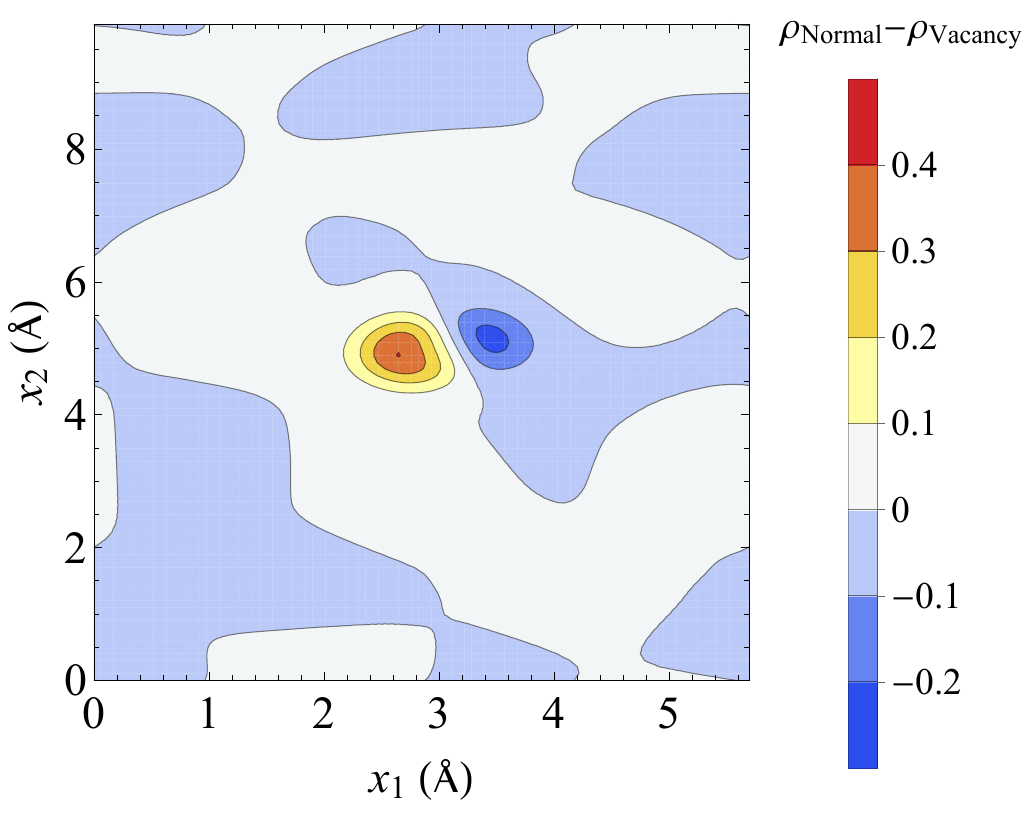}
\label{fig:CD111}}
\subfigure[]{
\includegraphics[scale=0.4]{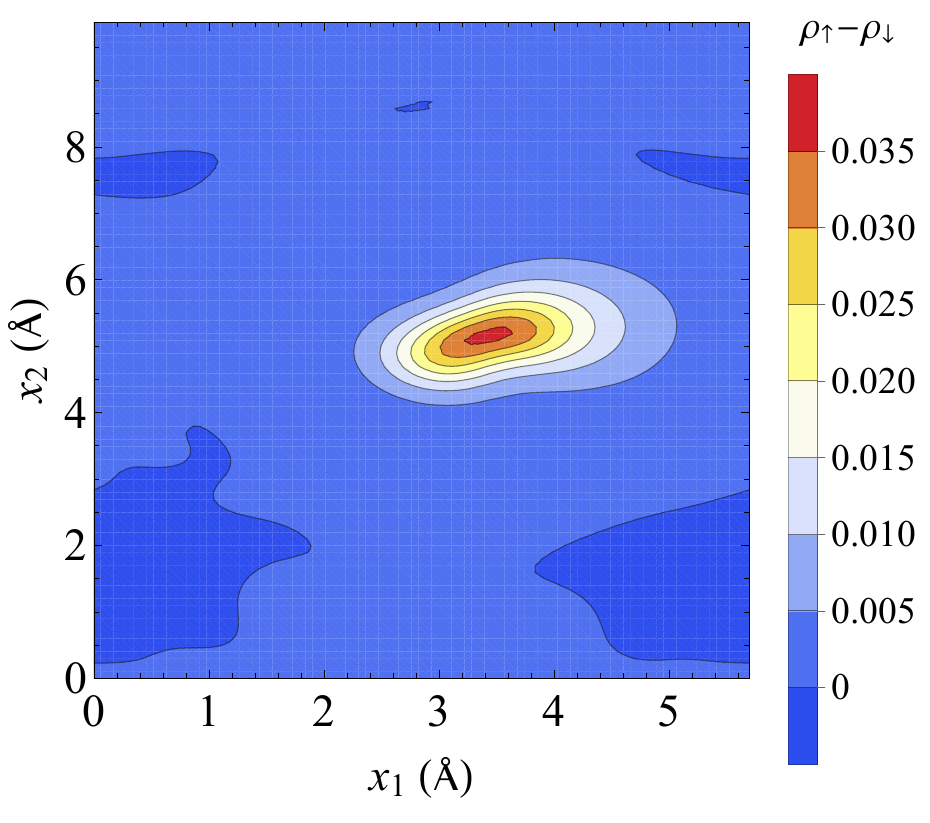}
\label{fig:SP111}}
\caption{(a) Supercell used in simulating 111 surface. (b) Difference in charge density on the surface for the supercell with an oxygen vacancy ($\rho_{vacancy}$) and without ($\rho_{normal}$). Unlike the 001 surface, the oxygen vacancy is offset from the plane containing the Ta atoms, removing the large differential seen in Fig. \eqref{001SurfaceDFT}. (c) Spin polarization on the surface of the supercell with an oxygen vacancy. Any significant contribution to spin-polarization is localized to the Ta atoms.}
\label{111SurfaceDFT}
\end{figure*}

\paragraph*{Emergent Magnetism: The (111) Surface}
To study the emergent surface magnetism on the (111) surface, we construct a supercell with the $c$-axis along the [111] direction. It is further constructed such that the surface terminations are equivalent and there exists two Ta atoms on the surface. Again 20 \AA of vacuum is added along the $c$-direction to limit finite size effects. The supercell is visible in Fig. \ref{111SurfaceDFT} (a). The relaxation procedure listed above for a single oxygen vacancy on the surface is repeated using identical energy cutoffs and k-point grids. The simulation estimates a total magnetic moment of $~6.67$ Bohr magneton for the full supercell with two surface Ta atoms. Importantly, the total magnetic moment has increased significantly from the (001) surface. We should note that the supercell used to model the (111) surface has 71 atoms as opposed to 35 atoms for the (001) surface, however, this still represents a significant qualitative increase and is in accordance with an increased zero-field charge Hall conductivity on the (111) surface. The process was repeated with no oxygen vacancies to ensure the system converged to a non-magnetic state.

The partial charges of each atom were then compared for the systems with and without the charge vacancy. The atom which displays most significant change in partial charge is the Ta atom on the surface, closest to the oxygen vacancy with a change in charge of approximately $0.5 e^{-}$. This is visible in the density plot shown in Fig. \ref{111SurfaceDFT} (b).  

\bibliography{KTO_bib}